\documentclass[twocolumn,prd,nofootinbib,superscriptaddress,eqsecnum,tightenlines,8pt]{revtex4}

\usepackage{fancyhdr}
\usepackage{amsmath,amsfonts,amssymb}
\usepackage[colorlinks,citecolor=blue,linkcolor=blue,urlcolor=blue]{hyperref}

\usepackage{color}
\definecolor{red}{rgb}{1,0,0}
\definecolor{gre}{rgb}{0,0.6,0}
\definecolor{blu}{rgb}{0,0,1}

\usepackage[pdftex]{graphicx}
\usepackage{mathrsfs}
\DeclareGraphicsExtensions{.jpg, .JPG , .png , .pdf, .gif}
\def\be{\begin{equation}}
\def\ee{\end{equation}}

\newcommand{\dd}{{\mathrm{d}}}
\newcommand{\OM}{{\mathbf{\Omega}}}

\begin{document}

\title{Cosmology without time:\\
What to do with a possible signature change from quantum gravitational origin?}

\author{Aur\'elien Barrau}
\email{Aurelien.Barrau@cern.ch}
\affiliation{
Laboratoire de Physique Subatomique et de Cosmologie, Universit\'e Grenoble-Alpes, CNRS-IN2P3\\
53,avenue des Martyrs, 38026 Grenoble cedex, France\\
}%

\author{Julien Grain}
\email{julien.grain@ias.u-psud.fr}
\affiliation{
Institut d'Astrophysique Spatiale, UMR8617, CNRS, Orsay, France, F-91405\\
Universit\'e Paris-Sud 11, Orsay, France, F-91405\\
}%


\begin{abstract}
Within some approaches to loop quantum cosmology, the existence of an Euclidean phase at high density has been suggested. In this article, we try to explain clearly what are the observable consequences of this possible disappearance of time. Depending on whether it is a real fundamental effect or just an instability in the equation of motion, we show that very different conclusions should be drawn. We finally mention some possible consequences of this phenomenon in the black hole sector.
\end{abstract}


\maketitle







\section{Introduction}
\label{sec:intro}
As a theory of the dynamics of space-time, General Relativity (GR) is most commonly phrased in a Lagrangian framework, where the gravitational fields is described by the metric tensor of (pseudo)riemanian manifolds. In this description, the "time" part of space-time is easily figured out from the signature of the metric tensor. Basically, Einstein's field equations are the equations of motion of a metric tensor which has the Lorentzian signature, $(-,+,+,+)$, and physical space-times are thus in the set of four dimensional, pseudoriemanian manifolds\footnote{The prefix "pseudo" is used to denote Lorentzian signature.}. The fact that a minus sign in the signature traces the presence of a "time" part in physical space-times can be intuitively understood in a causal sense. Because of that minus sign, the interval separating two space-time events can be either space-like (i.e. positive-valued),  light-like (i.e. null), or time-like (i.e. negative-valued). Two events which are separated by a space-like interval are causally disconnected since no geodesic traced by usual matter and radiation fields can link them. Roughly said, the distance between these two space-time events is such that no physical fields travels fast enough to cover that distance within the time interval separating the two events. For light-like and time-like separations however, physical fields can join the two events. Clearly, if the signature is Euclidean, $(+,+,+,+)$, such an interpretation is not possible since intervals are all positive-valued. \\

In the Hamiltonian language of GR, the signature of space-time is traced through the so-called algebra of hypersurface deformation. The Hamiltonian formulation of GR, as firstly proposed by Arnowitt, Deser and Misner \cite{PhysRev.116.1322}, relies on the foliation of 4-dimensional (pseudo)riemanian manifolds into a set of 3-dimensional hypersurfaces, $\left(\Sigma_t\right)_{t\in\mathbb{R}}$ (see \cite{lqg6} for details). On a given hypersurface, the canonical variables are the induced metric, $q_{ab}$, and its canonically conjugate momentum, $P^{ab}:=\frac{\delta S}{\delta \dot{q}_{ab}}$, with $\dot{q}_{ab}$ the derivative with respect to $t$ of the induced metric. In such a setting, the dynamics of the gravitational fields is the evolution of this set of canonical variables from one hypersurface to another. This phase space has however to be extended to include the lapse function, $N$, the shift vector, $N^a$, and their associated momenta, dubbed $C=\frac{\delta S}{\delta \dot{N}}$ and $C_a=\frac{\delta S}{\delta \dot{N}^a}$ respectively. This describes the adopted foliation, and since GR is independent of such a choice through diffeomorphism invariance, their is no evolution of $N$ and $N^a$, i.e. the Einstein-Hilbert action does not depend on $\dot{N}$ and $\dot{N}^a$. As a result, the momenta associated to $N$ and $N^a$ are constrained to zero and this should be preserved through evolution. This in turn shows that $N$ and $N^a$ are mere Lagrange multipliers and the Hamiltonian reads $\mathcal{H}=\int\dd^3x\left[NH(q_{ab},P^{ab})+N^aH_a(q_{ab},P^{ab})\right]$ ;  $H$ and $H_a$ are the Hamiltonian and spatial diffeomorphism constraints. Because this holds for any choice of the lapse function and the shift vector, $H\approx0$ and $H_a\approx0$ on solutions of GR, meaning that GR is a totally constrained system as a result of diffeomorphism invariance.  This property of the theory can be phrased into an algebraic structure. One can define the smeared constrained, $S(N)=\int\dd^3xNH$ and $D(N^a)=\int\dd^3xN^aH_a$, and their time evolution is given by the Poisson brackets $\left\{\mathcal{H},S\right\}$ and $\left\{\mathcal{H},D\right\}$. Solutions of GR are such that $H=H_a=0$ and this should obviously be preserved across evolution.  This can be shown to be equivalent to an algebraic structure satisfying the constrained themselves:
\begin{align}
	&\left\{D(M^a),D(N^a)\right\}=D\left(M^b\partial_bN^a-N^b\partial_bM^a\right), \label{eq:algebra1} \\
	&\left\{D(M^a),S(N)\right\}=S\left(M^b\partial_bN-N\partial_bM^b\right), \label{eq:algebra2}  \\
	&\left\{S(M),S(N)\right\}=sD\left(q^{ab}(M\partial_bN-N\partial_bM)\right),\label{eq:algebra3} 
\end{align}
with $s=1$ for the Lorentzian signature, and $s=-1$ for the Euclidean one \cite{lqg6}. This means that the subspace of the phase space which satisfies the constraints, called the surface of constraints, is preserved through evolution. The constraints, also denoted $\mathcal{C}_I$ with $\mathcal{C}_1\equiv D$ and $\mathcal{C}_2\equiv S$, are said to form a system of first-class, i.e. $\left\{\mathcal{C}_I,\mathcal{C}_J\right\}={f^K}_{IJ}(q_{ab},P^{ab}) \mathcal{C}_K$ with ${f^K}_{IJ}$ structure functions depending on the phase-space variables.

Interestingly enough, the signature explicitly appears in the last Poisson bracket of the algebra of constraints. This is in fact not a surprise since the above Hamiltonian description admits a clear geometrical interpretation \cite{HOJMAN197688}. Solution of GR are 4-dimensional (pseudo)riemanian manifolds. In the Hamiltonian framework, one embeds hypersurfaces in this 4-dimensional structure. The embedding defines two possible way of deforming the embedded hypersurfaces: one can either displace a point {\it within} the hypersurface, or one can displace a point in the direction orthogonal to the considered hypersurface. These two transformations are described by generators, $\mathcal{D}_a$ and $\mathcal{D}$, which respectively generates diffeomorphisms preserving $\Sigma_t$, and diffeomorphisms orthogonal to $\Sigma_t$. The two generators form an algebraic structure called the {\it algebra of hypersurface deformation} that is exactly the above algebra formed by the constraints themselves. In other words, the Hamiltonian and spatial diffeormorphism constraints are {\it representations} of the transformations of the embedded hypersurfaces. And, in fact, this can even be used in a way to construct GR \cite{HOJMAN197688}. Since evolution of any phase space variables representing a given physical field as e.g. $(q_{ab},P^{ab})$ for gravitational degrees of freedom (but this applies to any fields), is no more than transporting this variables from one hypersurface to another, the functional which generates evolution should be representations of the generators of the hypersurface deformation. Obviously, the algebra of hypersurface deformation depends on the signature of the original 4-dimensional manifold in which hypersurfaces are embedded. And this is why the algebra of constraints, being a representation of the hypersurface deformation, traces the signature of the original 4-dimensional structure. Following this interpretation and the constructive approach of \cite{HOJMAN197688}, the algebra of constraints, Eqs. (\ref{eq:algebra1}) to (\ref{eq:algebra3}), is a consequence of the algebra of the hypersurface deformation. \\

Over the last decade, results in loop quantum cosmology (LQC hereafter) \cite{Barrau:2013ula}, also in the case of spherically symmetric space-times or  Gowdy space-times \cite{Bojowald:2015zha,Bojowald:2015sta}, surprisingly show that at an effective level, the algebra of {\it quantum-corrected constraints} might be deformed as compared to the one obtained in GR \cite{eucl3}. More specifically in the so-called {\it deformed-algebra} approach of LQC, it can be shown that to preserve an algebraic structure, the last Poisson bracket should be modified to $\left\{S(M),S(N)\right\}=D\left(\OM  q^{ab}(M\partial_bN-N\partial_bM)\right)$ where $\OM$ is a function of the gravitational phase-space variables. Remarkably enough, this function can even change its sign in some concrete cases: it is positive-valued well below the Planck scale but becomes {\it negative-valued} close to the Planck scale. If the sign in front of the left-hand-side of the Poisson bracket in Eq. (\ref{eq:algebra3}) is to be interpreted as the signature of the underlying 4-dimensional structure, then this would mean that at the Planck scale, space-time becomes Euclidean, or stated otherwise, that time has disappeared. 

Obviously the full picture is more subtle than that. In GR, we know the underlying 4-dimensional manifold to be pseudoriemanian, and the equivalence between the algebra of hypersurface deformation and the algebra of constraints is clearly established. In LQC however, the deformation of the last Poisson bracket is obtained at the level of the algebra of constraints. Whether this traces or not a deformation of an underlying, signature-changing, 4-dimensional space in which time may disappear, is still highly debated. (What is only certain is that if such a structure does exist, this cannot be a pseudoriemanian space.) More specifically in LQC, the presence of $\OM$ in the algebra of constraints could be interpreted as a mere instability of fields propagating on a quantum space-time, and not as a true signature change of the quantum-corrected space-time. The good point however is that depending on the adopted interpretation, this leads to different observational consequences in the amplitude of cosmological inhomogeneities produced in the early Universe, which in principle open the window for observational tests of this remarkable feature. \\

This article aims at presenting the possible observational consequences of signature change as obtained in LQC. Loop quantum gravity is a well defined theory (see, {\it e.g.} \cite{lqg3} for a review). It is based on Ashtekar's formulation of GR and can be derived from different approaches: canonical quantization of GR, covariant quantization of GR, and quantization of geometry. There are interesting and promising attempts to derive the cosmological dynamics from the full theory (see, {\it e.g.}, \cite{Bianchi:2010zs}). However most of the results obtained in LQC, in particular when dealing with perturbations, are based either on effective equations or on heavy hypotheses. A possible way to address this question is to try to deal with quantum fields on a quantum background, as done in the {\it dressed metric} approach (see, {\it e.g.}, \cite{agullo2}). Another way to face the problem of perturbations is to put the emphasis on the consistency of the effective equations, as done in the {\it deformed-algebra} approach (see, {\it e.g.}, \cite{eucl3}). 

Here, we mainly -- but not only -- focus on the {\it deformed-algebra} approach which can also be seen as a kind of embedding of GR in a more general framework. In the first part, we recall the reasons why a change of signature can be expected in LQC. In the second section, we explain the consequences of this phenomenon if it is understood as an effective instability in the equation of motion of modes in Fourier space. In the third section, we focus on the consequences if the phenomenon is understood as a real change of signature at the deepest level. Then, we explore in the next section some possible consequences for black holes. Finally, we briefly discuss some open questions related to the signature change in the last section, and which might be of phenomenological relevance.


\section{Origin of the change of signature in loop quantum cosmology}
\label{sec:lqc}
\subsection{Classical description in a nutshell}
Our Universe is commonly described by a {\it perturbed} Friedmann-Lema\^\i{tre}-Robertson-Walker (FLRW) metric, introducing the line element 
\begin{eqnarray}
	\dd s^2&=&a^2(\eta)\big[-(1+A)\dd\eta^2+2B_a\dd x^a\dd\eta\nonumber\\
	&+&\left(\gamma_{ab}+E_{ab}\right)\dd x^a\dd x^b\big], \label{eq:ds}
\end{eqnarray}
with $\eta$ the conformal time, $a(\eta)$ the scale factor, and $\gamma_{ab}$ the flat, 3-dimensional euclidean metric (note that the perturbed FLRW space-time is not the FLRW space-time). The background evolution is given by the evolution of $a(\eta)$, and $A,~B_a$ and $E_{ab}$ are perturbative degrees of freedom describing cosmic inhomogeneities. The latter are functions of space and time, $(\eta,\vec{x})$, and they are usually decomposed into scalar, vector and tensor modes. Similarly, the matter content of the Universe is decomposed into an isotropic and homogeneous part, plus some perturbations. Because of the diffeomorphism invariance of the theory, it is possible to introduce changes of the coordinate system on the strictly FLRW space-time which mimics perturbative degrees of freedom. The true perturbations are thus identified by searching for gauge-invariant quantities, that is combinations of the scalar, vector, and tensor decompositions which are invariant under gauge transformations, $x^\mu\to x^\mu+\epsilon^\mu(x^\mu)$.

Loop quantum cosmology is phrased in the language of the Ashtekar formalism of general relativity, and the classical theory of cosmological perturbations can be recovered in this language. In this frame, the perturbative degrees of freedom are introduced as follows (the background, isotropic and homogeneous degrees of freedom are denoted as barred quantities): the Ashtekar connection reads $A^i_a=\gamma\bar{k}\delta^i_A+\delta A^i_a$, the densizied triad reads $E^a_i=\bar{p}\delta^a_i+\delta E^a_i$, and the lapse function and the shift vector respectively read $N=\bar{N}(1+\phi)$ and $N^a=\bar{N}^a+\partial^aB$. (Note that $\bar{p}\equiv a$ and one can choose $\bar{N}=a$ to work with conformal time ; $\bar{N}^a=0$ as a result of homogenity and isotropy ; $\gamma$ is the Barbero-Immirzi parameter\footnote{Note that this parameter is irrelevant at the classical level.}.) A similar decomposition can be performed for the matter content, which for a scalar field is $\varphi=\bar{\varphi}+\delta\varphi$ and $\pi=\bar{\pi}+\delta\pi$. The dynamics is thus given by the Hamiltonian linearized at second order, i.e.
\begin{eqnarray}
	S[N]&=&\displaystyle\int_{\Sigma_t}d^3x\left[\bar{N}\left(H^{(0)}+H^{(2)}\right)+\delta N H^{(1)}\right], \label{eq:scl}\\
	D[N^a]&=&\displaystyle\int_{\Sigma_t}d^3x\left[\bar{N}^a\left(D^{(0)}_a+D^{(2)}_a\right)+\delta N^a D^{(1)}_a\right]. \label{eq:dcl}
\end{eqnarray}
The explicit expressions, as well as the Scalar-Vector-Tensor decomposition in the Ashtekar formalism can be found in \cite{eucl2} (note that the above constraints contain both the gravitational sector and the matter sector). The dynamics generated by the zeroth order just give the Friedmann equations. Since $S$ and $D$ generate time and space diffeomorphisms, the first order of the linearized constraints are used to build the gauge invariant perturbative degrees of freedom as the combination which has a vanishing Poisson bracket with $S^{(1)}$ and $D^{(1)}$ \cite{Langlois:1994ec,Cailleteau:2011mi}. The dynamics of the perturbations evolving in a curved background whose dynamics is given by the Friedmann equations\footnote{The dynamical variables in $S^{(2)}$ and $D^{(2)}$ are the perturbative degrees of freedom. However, this two functions are also "parametrized" by some background variables which encode the impact of the background variables on the perturbative ones.}, is finally obtained from their Poisson bracket with $S^{(2)}$ and $D^{(2)}$ (note that by construction, the gauge-invariant perturbative degrees of freedom have a vanishing Poisson bracket with the first order constraints). 

As a result of this process, one obtains the equation of motion of the so-called Mukhanov-Sasaki variables for the perturbative degrees of freedom. In the case of a scalar field as the matter content, there are only scalar and tensor perturbations. As an example, the equation of motion of the tensor modes in spatial Fourier space is
\begin{equation}
	\left(ah_k\right)''+\left(k^2-\frac{a''}{a}\right)\left(ah_k\right)=0,
\end{equation}
where $h_k$ and $k$ are the amplitude and the wavenumber of the gravitational waves, and $(a''/a)$ encodes the impact of the background curvature on the propagation of the tensor modes.

Finally, one should note that though the constraints in Eq. (\ref{eq:scl}) \& (\ref{eq:dcl}) are truncated constraints, they still satisfy the classical algebra of hypersurface deformation.

\subsection{Loop quantum cosmology at an effective level}
In LQC, the dynamics of our Universe should be described by a quantum space-time, at least in the Planck era, including {\it a priori} cosmic inhomogeneities as perturbative degrees of freedom. There is a consensus at the background level (see e.g. \cite{lqc9,lqc10} for reviews), but different approaches have been developed if cosmological perturbations are to be accounted for.  Here we will focus on the so-called deformed-algebra approach, which precisely leads to a deformation of the algebra of constraints. (We refer to \cite{Grain:2016jlq} for a more detailed overview, to e.g. \cite{agullo1,agullo2,agullo3} for alternatives, and to \cite{Barrau:2016nwy} for conceptual discussions of the different approaches.) \\

The deformed-algebra approach works at an effective level by adding quantum corrections to the truncated constraints. The idea is to consider that cosmological perturbations evolve on a background quantized using loop quantization techniques. In LQC, this is not the connection itself which is quantized but its holonomy. Working at an effective level means that the quantum nature of the background is modeled as modifications of the classical, truncated constraints. The Poisson bracket is however undeformed. The intent is to consider perturbations on a quantum background, and only background variables entering the truncated constraints will be modified (keeping in mind however that background variables are involved in the first and second order of the constraints). The modifications should at some point capture the fact that holonomies of the connection are used. This means that in the truncated, classical constraints, the connection $\bar{k}$ is replaced by $f(\bar{k})$ where $f$ is a pseudoperiodic function in line with the idea of quantizing holonomies of the connection. We stress that holonomy corrections are introduced order by order allowing for different $f_i$ functions, i.e. there is no a priori reason for $f_0(\bar{k})$ in $H^{(0)}$ to be the same than $f_2(\bar{k})$ introduced in $H^{(2)}$\footnote{This may appear as odd. However, these modifications are supposed to account for quantum corrections. One may thus imagine that these differences trace some operator ordering which differs order by order at the quantum level, though there is no formal proof of this.}.

There are two important points to underline about this approach. First of all, this procedure is performed prior to the fixing of any gauge-invariant variables for the perturbations. The reason is that the Hamiltonian and diffeomorphism constraints generate both the dynamics of the background {\it and} its gauge transformations. If its dynamics is quantum corrected, one can expect the gauge transformations to be modified too, and one should thus not fix a priori the gauge-invariant, perturbative degrees of freedom. Second, it was rapidly realized that among the many possible choices for the $f$'s, many lead to truncated, quantum-corrected constraints that do not form a closed algebra anymore, i.e. $\left\{\mathcal{C}^{QC}_I,\mathcal{C}^{QC}_J\right\}={f^K}_{IJ}(q_{ab},P^{ab}) \mathcal{C}^{QC}_K+\mathcal{A}_{IJ}$ (the superscript $QC$ means "quantum-corrected at an effective level", and $\mathcal{A}_{IJ}$ are anomalous terms). As a consequence, and taking into account lessons from GR, the choice of the $f$ functions is made such that the algebra of the truncated, quantum-corrected constraints is closed (with the inclusion of counterterms). However to achieve such a goal, one should allows for a possible modification of the structure functions of the algebra, i.e. ${f^K}_{IJ}(q_{ab},P^{ab})\to {}_{QC}{f^K}_{IJ}(q_{ab},P^{ab})$. These two "remarks" are keypoints of the overall approach: if gauge-invariance is to be taken seriously at an {\it effective level} of the {\it truncated constraints}, then it is mandatory to allow for the structure functions of the algebra to be modified, which can potentially lead to a signature change if the third quantum-corrected, structure functions, ${}_{QC}{f^1}_{22}$, becomes negative-valued. \\

At the zeroth order, i.e. background only, the full quantization {\it \`a la} loop of this minisuperspace has been performed, which determines the choice of $f$ at this order. Taking the semi-classical limit of the quantum constraints shows that an effective description is obtained by replacing $\bar{k}$ with $\sin(\gamma\mu\bar{k})/(\gamma\bar\mu)$, where $\bar\mu=\sqrt{\Delta/\bar{p}}$ and $\Delta=2\sqrt{\pi}\ell^2_\mathrm{Pl}\gamma$ is the minimal area gap as obtained from loop quantization. The minimal area gap defines the maximal energy density in the modified Friedmann equation at which the bounce occurs, $\rho_c=3/(8\pi G \gamma^2\Delta)$. At this level, the algebra of constraints (thus truncated at the zeroth order) is unmodified. However, at this level one cannot have access to ${}_{QC}{f^1}_{22}$. This is because this structure function is multiplied by the quantum-corrected diffeomorphism constraints which is vanishing at the background level because of exact homogeneity and isotropy. 

At first and second order in the constraints, the $f$ functions are built from pseudoperiodic functions, $\bar{k}\to\sin(n\gamma\bar{\mu}\bar{k})/n\bar{\mu}\gamma$ with $n\in\mathbb{N}$, and one also allows for additional counterterms to cancel the anomalous term of the algebra (note that the counterterms are constrained to vanish in the classical limit). The full calculations of the different Poisson brackets of the constraints, and how the integer $n$ and the counterterms should be chosen to finally obtain a closed algebra are presented in \cite{tom1,tom2}. Despite tedious calculations, the solution with holonomy correction is {\it unique} (under reasonable assumptions), and surprisingly simple in terms of the algebra of the constraints, i.e.:
\begin{align}
	&\left\{D^{QC}[M^a],D^{QC}[N^a]\right\}=D^{QC}\left(M^b\partial_bN^a-N^b\partial_bM^a\right), \\
	&\left\{D^{QC}[M^a],S^{QC}[N]\right\}=S^{QC}\left(M^b\partial_bN-N\partial_bM^b\right), \\
	&\left\{S^{QC}[M],S^{QC}[N]\right\}=D^{QC}\left(\OM(\bar{k}) q^{ab}(M\partial_bN-N\partial_bM)\right),	
\end{align}
with $\OM(\bar{k})=\cos(2\gamma\bar\mu\bar{k})=1-2\rho/\rho_c$, which only depends on the background variables. We stress that for $\rho>\rho_c/2$ (that is close to the bounce), $\mathbf{\Omega}$ becomes negative-valued. Here, the algebra of constraints is not only  deformed, but this deformation can potentially lead to a change of signature, as explained in our introduction.\\

The above-described path to the deformed-algebra approach assumes that one performs first a linearization of the Hamiltonian for the FLRW space-time, and, then, implements order by order the quantum corrections at an effective level. A different strategy was adopted in \cite{WilsonEwing:2011es}:  the holonomy corrections are introduced at an effective level first, and then a linearization of the effective FLRW space-time is performed. To do this, the Hamiltonian and diffeomorphism constraints are written in terms of the line element Eq. (\ref{eq:ds}) {\it without} performing the linearization in $A,~B_a$ and $E_{ab}$. Then, working in the longitudinal gauge and restricting the study to scalar inhomogeneities (which is in fact sufficient to make the deformation of the algebra explicit \cite{eucl2}), it is possible to introduce holonomy corrections at an effective level, and subsequently to perform the linearization of the quantum-corrected constraints. This path also leads to the same deformation of the algebra. As compared to the approach developed in \cite{eucl2,tom1,tom2} and described above, this path is restricted to a specific gauge. However, modifying the constraints in such a way that their algebra is deformed, is needed prior to linearization. This suggests that the deformation of the algebra might not be an artifact of the perturbative expansion (see also \cite{eucl3} for additional arguments on this point).

Alternatively, a more quantum-setteled approach was developed in \cite{ed}. The idea is to model inhomogeneities at a quantum level by allowing for the lattices to slightly differ from one to each other, still working in the longitudinal gauge. The commutators between the quantum constraints of this model are explicitly computed, and shown to give the above deformed algebra at an effective level by taking both the classical limit and the continuum limit of the commutators. Again, this path is restricted to a specific choice of gauge for the inhomogeneities. However, the algebra of the constraints is here demonstrated at a ``more quantum" level. \\

A final remark is in order. The different paths to account for cosmological perturbations propagating in a quantum background presented here all lead to a deformation of the algebra of constraints. However, none of them were firstly built with that purpose, but instead with the aim of deriving consistent equations of motion for perturbations. The fundamental requirement is thus not to have a deformed-algebra, but to make sure that the {\it effective} theory is anomaly-free. The former is a consequence of the latter, and not vice-versa ; and if a negative-valued $\OM$ is to be really interpreted as a signature change, this would be the consequence of imposing covariance at an effective level. 

Obviously, none of these approaches are free of hypotheses, and none of them is deduced from the full loop quantum gravity theory (in that respect, alternative approaches, as already mentioned, are also welcome). As an example, the approach of \cite{eucl2,tom1,tom2} does not assume any gauges but starts from a perturbative treatment of the total Hamiltonian, while the approach of \cite{WilsonEwing:2011es} does not assume linearization but works in a specific gauge. There are however different paths for a treatment of cosmic inhomogeneities in a loop quantum universe that converge towards such a deformation of the algebra of constraints.


\subsection{Equation of motion for perturbations}
The three different paths presented above show that the algebra of constraints could be deformed once loop quantum corrections are taken into account, at least at an effective level and in the context of cosmological perturbations. This is nevertheless not sufficient to bridge contact with astronomical observations. 

Fortunately, the procedure described above does not only lead to the quantum-corrected algebra of constraints. In fact, its first purpose is to provide the effective equations of motion for the perturbations propagating in a quantum background, and thus to bridge contact with astronomical observations of cosmic inhomogeneities. The overall procedure univocally determines the quantum-corrected constraints up to the second order (at least with holonomy corrections in the cosmological context) from which one can derive the equations of motion for the perturbations.  \\

The way of proceeding is identical to the classical treatment of cosmological perturbations. From the quantum-corrected first order constraints, one defines the gauge-transformations and thus the gauge-invariant variables for perturbations as the ones having a vanishing Poisson bracket with $H^{(1)}_{QC}$ and $D^{(1)}_{QC}$. In particular, one can show that for a scalar field as the only matter content, there are only scalar and tensor perturbations, and among the different possible choices of gauge-invariant variables, it is possible to construct the analog of the Mukhanov-Sasaki variables \cite{Cailleteau:2011mi}. Their equation of motion is then obtained from their Poisson brackets with the second-order, quantum-corrected constraints. Focusing on the case of the tensor modes, the equation of motion is
\begin{equation}
	\left(z_\mathrm{T}h_k\right)''+\left(\OM k^2-\frac{z''_\mathrm{T}}{z_\mathrm{T}}\right)\left(z_\mathrm{T}h_k\right)=0, \label{eq:eomtensor}
\end{equation}
with $z_\mathrm{T}=a/\sqrt{\left|\OM\right|}$. The deformation of the algebra appears twice: in the function ${z}_\mathrm{T}(\eta)$ which measures the impact of the curved, and now quantized, background, and in front of $k^2$. A similar equation is obtained for the scalar modes \cite{eucl2}. We stress that in the classical limit ($\OM\to1$), the classical equations of motion are recovered.

\section{Treatment as an instability}
\label{sec:instability}
Focusing on the equations of motion of cosmological perturbations, the appearance of the $\OM$ term can be simply interpreted as a instability. This is  possible if initial conditions are set during the classical contraction. In LQC, the dynamics is indeed such that the background goes through a phase of classical contraction prior to the quantum bounce, and smoothly transits to a phase of classical expansion through the regular, quantum bounce. During most of the contraction and of the expansion, $\OM>0$ since $\rho<\rho_c/2$, and the equation of motion for perturbations is hyperbolic. Only around the bounce is $\OM$ negative-valued (as a result of $\rho_c/2<\rho\leq\rho_c$) leading to an elliptic equation of motion. This means that if initial conditions are set during the contraction phase\footnote{The initial value of $h_k$ and its first time-derivative $h'_k$ are fixed on a constant-time hypersurface, $\eta=cste$, in the asymptotic past of the contraction.}, one can solve the system for an {\it initial value} problem since the final predicted outputs are the primordial power spectra during the classical expansion following the bounce. In both regions, the equations of motion are hyperbolic. Moreover, the initial state can be chosen as the simple Minkowski vacuum in the asymptotic past of the classical contraction where curvature is negligible. However, the cosmological perturbations experience a phase of instability around the bounce because $\OM<0$ in this region. (This viewpoint can be more intuitively understood in analogy with density waves propagating in a fluid whose flow is first subsonic, then supersonic, and then subsonic again. An instability will appear during the supersonic era of the flow.)

This interpretation as an instability imposes to set the initial conditions in the remote past of the contracting phase. Clearly, at the bounce where the equation of motion is elliptic, one cannot solve for an initial value problem, and there is in addition no clear meaning of what would be a possible vacuum state for perturbations in this region. \\

The observational consequences of this scenario have been exhaustively investigated considering a massive scalar field as the matter content of the universe \cite{lcbg,Bolliet:2015bka,Schander:2015eja,Bolliet:2015raa}. With this specific matter content, the universe additionally goes through a phase of inflation shortly after the bounce, and the primordial power spectra for scalar perturbations \cite{Schander:2015eja}, and tensor perturbations \cite{lcbg,Bolliet:2015bka}, are thus evaluated at the end of inflation. The instability due to $\OM$ yields a clear feature in the small-scale part of the primordial power spectra: for scales such as $k\gtrsim\sqrt{24\pi G\rho_c}$, the primordial power spectra oscillate with an exponentially raising envelop, i.e. $\mathcal{P}_{\mathrm{S(T)}}\propto\exp\left(k/\sqrt{6\pi G\rho_c}\right)$. Without the instability, the predicted power spectra at small scales would be identical to the standard prediction of inflation \cite{Bolliet:2015bka}. For smaller values of $k$ (i.e. larger scales), the predicted power spectra differ from the standard prediction of inflation because of the contraction preceding the bounce. This subtle effect boosts the power. However, it does not differ from the primordial power spectra as predicted in a bouncing universe {\it without} the $\OM$ correction \cite{Bolliet:2015bka}.

This obviously raises questions, in particular regarding the validity of the perturbative approach or regarding the regularization of the energy of such a state. It was however shown that if this result is taken as it is, it is in all cases in contradiction with astronomical observations \cite{Bolliet:2015raa}. The reason is simple: the statistical properties of the cosmic microwave background anisotropies which result from the primordial power spectra derived in a bouncing universe {\it with} the instability, drastically differ from the observed ones that favor a nearly-scale invariant power spectrum for scalar perturbations ; and this is so irrespectively of the range of scales in the primordial power spectra that are probed by the cosmic microwave background anisotropies \cite{Bolliet:2015raa}. It is  a very nice result that a reasonable approach to quantum gravity with the correct IR limit and predicting inflation is exclude by data: this shows that quantum cosmology is non-trivially falsifiable !

\section{Treatment as a real disappearance of time}
\label{sec:silent}
In the previous section, modes were propagated through the bounce as if nothing deeply specific was happening. However, if the change of sign of the $\OM$ factor is to be understood as a disappearance of time at the deepest level, it is hard to give any sense to a propagation through the bounce. Even the Fourier transformation over a basis of plane waves is not anymore well defined. This, however, does not mean that nothing at all can be said. There are at least two possible ways to face this new situation. It is important to stress that, as explained in  \cite{eucl3}, there are indications that this effect might be real at the deepest level: perturbations are here just used as a trace field to show the structure of spacetime and it might very well be that the deformed algebra holds also at the non-perturbative level. It cannot be seen by studying the background alone as the relevant Poisson bracket  identically vanishes.

\subsection{Focus on the silent surface}
The first possible view consists in focusing on the silent surface, {\it i.e.} the $\OM=0$ surface, as the correct place where to put initial conditions. This has been studied in \cite{Mielczarek:2014kea}. The key question is obviously  the one related with the correct choice of the vacuum. Interestingly it is possible to make assumptions directly about the spectrum itself and not only about mode functions. The nonlinear equation of motion for the spectrum is
 \begin{equation}
\frac{\dd^2 \mathcal{P}}{\dd\eta^2}-\frac{1}{2\mathcal{P}} \left( \frac{\dd \mathcal{P}}{\dd\eta} \right)^2
+2\OM k^2 \mathcal{P}+2 \frac{\dd\mathcal{P}}{\dd\eta} \frac{z'}{z}
-\frac{1}{2z^4 \mathcal{P}} \left( \frac{k^3}{2\pi^2} \right)^2=0.   
\label{Ermakow}
\end{equation}
This can be also written as a set of two first order differential equations. The Wronskian condition together with heuristic probabilistic arguments can be used to determine the free parameters entering the solutions of this differential equation.  As shown in \cite{Mielczarek:2013xaa}, the equation of motion of tensor modes can be recovered by considering a wave equation on the effective metric 
\begin{equation}
g^{eff}_{\mu\nu}\dd x^{\mu} \dd x^{\nu} =  -\sqrt{\OM} a^2 \dd\eta^2
+\frac{a^2}{\sqrt{\OM}} \gamma_{ab}\dd x^a\dd x^b.  \label{effectivmetric}
\end{equation}
One can then define a Hamiltonian, quantize it and show that the vacuum expectation value is
\begin{equation}
\langle 0 | \hat{H} | 0 \rangle = \delta^{(3)}(0) \frac{1}{2}\int \dd^3k E_k, 
\end{equation}
with $E_k = |f_k'|^2+\omega_k^2|f_k|^2$ and $\omega_k^2 = \OM k^2 -\frac{z^{''}}{z}$. The energy can be minimized if and only if $\omega_k^2>0$. Interestingly, in the Lorentzian regime ($\OM>0$), it is always 
possible to find values of $k$ for which $\omega^2_k$ is positive. If $\frac{z^{''}}{z}$ is negative, this 
is the case for any $k$. If $\frac{z^{''}}{z}$ is positive, this requires sufficiently large (sub-Hubble) $k$-valued mode. But the situation
is different in the Euclidean regime where $\OM<0$: the positiveness of $\omega^2_k$
can be satisfied only if $\frac{z^{''}}{z}$ takes a negative value.

In the Lorentzian domain, the vacuum state is well defined at the sub-Hubble scales but this reverses in the Euclidean regime.
The fact that the vacuum state can be well defined at  super-Hubble scales is a new feature.  
This might be used to impose initial conditions for the super-Hubble 
modes at the beginning of the Lorentzian regime, where the vacuum sate is not usually well defined. In particular, 
one might expect that through a possible quantum tunneling transition from the Euclidean to the Lorentzian phase, the structure of the Euclidean vacuum at the super-Hubble scales might --at least partially-- define the configuration of the 
perturbations at the beginning of the Lorentzian regime.\\

The vacuum can be set either at $\OM=-1$, or at  $\OM>0$, or at $\OM \approx 0$. This latter case is the most interesting one. For a barotropic fluid with an equation of state $w=-1$, and for  positive values of $\omega_k^2$,
the mode functions for the state of vacuum can be shown to be
\begin{equation}
|f_k|^2 = \frac{1}{2  \sqrt{\OM\left(k^2- \frac{1}{3}\kappa\rho_c a^2\right)}}.
\end{equation}
The vacuum is then characterized by the  spectrum 
\begin{equation}
\mathcal{P}_{\phi}(k) = \frac{\sqrt{3} \sqrt{|\OM|}}{4 \pi^2 \sqrt{\kappa \rho_c }}  \left( \frac{k}{a}\right)^3 
\propto \sqrt{|\OM|} k^3,
\end{equation} 
which is a white noise spectrum modulated by $\OM$. There are no fluctuations strictly at the surface of silence but interesting features remain around the surface. 
The correlation function is  trivially vanishing exactly at $\OM=0$ but should take a non-zero value up to a scale $\xi$ near the surface. This can even be analytically invesitgated by considering a correlation function 
\begin{equation}
G(r) = \left\{ \begin{array}{ccc}    G_0 & \text{for}  & \xi \geq r \geq 0,  \\ 
0 & \text{for} & r>\xi.  \end{array} \right\}
\end{equation}
and taking the limite $k\xi \ll 1$. The related analytical results are given in \cite{Mielczarek:2014kea} and show that a potentially interesting phenomenology might emerge.\\

An important issue is the following. Except if the duration of inflation is very close to its minimum allowed value, at least some modes will become transplanckian before reaching the silence surface when going backward in time. A consistent way to take this into account is to notice that due to the deformation of the algebra of constraints, the Poincar\'e algebra should be deformed as well \cite{Bojowald:2012ux, Mielczarek:2013rva}. This can be modeled by considering a $k$-dependence in the $\OM$ function. A quite natural modification could be
\begin{equation}
\OM = \left(1- 2\frac{\rho}{\rho_{\text{c}}}\right)\left(1 - \frac{1}{m_{\text{Pl}}^2}  \left( \frac{k}{a} \right)^2\right). 
\label{BetaGeneralized}
\end{equation}  
Setting initial conditions according to the usual vacuum-type normalization of the modes, this leads to a nearly flat spectrum in the UV limit. On the other hand, the IR part, if visible because of a number of inflationary e-folds not too large, can be turned into a flat spectrum but this requires a quite high level of fine-tuning.  More precisely, the relevant modes must enters the Hubble horizon when $\omega=\omega_1>-1/3$ and exits the Hubble horizon when $\omega=\omega_2<-1/3$, where $\omega_1$ and $\omega_2$ fulfill
\begin{equation}
\frac{1}{1+3\omega_1}-\frac{1}{1+3\omega_2}=\frac{3}{4}.
\end{equation}

\subsection{The Tricomi problem}
Another interesting approach was considered in \cite{Bojowald:2015gra}. It basically treats carefully the fact that for $\OM<0$, the usually hyperbolic equation become elliptic. Partial differential equations of mixed type have been studied in mathematics for nearly a century. The simplest example is the co-called  Tricomi problem
\begin{equation}
\frac{\partial^2 u}{\partial y^2}+y\frac{\partial^2u}{\partial x^2}=0.
\label{TricomiEQ}
\end{equation}
This type of equation is an approximation to the propagation equation of LQC around one of the  boundaries of the elliptic regime, where $\OM=0$. 

Solutions to this equation are well defined and stable if
correct boundary conditions are imposed. Let us call $U_1$ the part of the elliptic region where the solution is looked for, and $U_2$ the equivalent part of the hyperbolic region. The key-point is that the initial values for $u$ are not specified in the same way in $U_1$ and $U_2$. When adapting the mathematical treatment of the Tricomi problem to the cosmological case, the resulting view is that this mixed-type partial differential equations lead to an interesting balance between deterministic cyclic models and singular Big-Bang models. Divergences are avoided, but initial data set in the remote past of the contracting branch do not fully determine all of the space-time structure and evolution. For each mode, a function has to be specified at the beginning of the expansion branch even if one has already chosen initial values for the contraction phase. The normal derivative of the field is not free and carries  information about the pre-Big Bang. Quite a lot is still to be done to draw clear conclusions from this approach.\\

This way of thinking might seem in contradiction with the previous one whereas the problem is mathematically well posed. The main reason for this apparent discrepancy is that this latter approach is constructed for the avoidance of instabilities whereas the previous one precisely considers the instable solution associated with the growing exponential. At this stage it is too early to decide which one is potentially correct.  

\section{Possible consequences for black holes}
\label{sec:bh}
If real in cosmology, it is reasonable to conjecture that this effect also occurs in black hole physics. This has not yet been studied except in \cite{Bojowald:2014zla} where it is stated that the information paradox is made worst by quantum gravity because the information is necessarily lost when it reaches the equivalent of the silence surface. This view should however be revised in the light of the new model of bouncing black holes recently proposed in \cite{Rovelli:2014cta,Haggard:2014rza} and whose  phenomenology was considered in \cite{Barrau:2014hda,Barrau:2014yka,Barrau:2015uca,Barrau:2016fcg}. The basic idea is that when matter or radiation reaches the Planck density, quantum gravity generates a high enough pressure to counterbalance the classically attractive gravitational force. For a black hole, this means that the collapse of matter should stop before the central singularity is formed. The standard event horizon of the black hole is replaced by a ``trapping'' horizon \cite{Ashtekar:2005cj} which is locally equivalent, but from which matter can eventually bounce out. It was  shown that a realistic effective metric satisfying Einstein equations everywhere outside the quantum region can be constructed. The model describes a tunneling from the classical in-falling black hole to the classical emerging white hole. Although very fast in proper time, the full process is seen in extremely slow motion from the outside because of the huge time dilatation inside the gravitational potential. In this approach, the Hawking evaporation can be neglected and considered as a small dissipative correction.\\

At the qualitative level it might be expected that if the  change of signature is seen as an effective phenomenon, the physical modes collapsing during this process will undergo the  same evolution that what happens in cosmology. This means that the bounce will not anymore be fully time-symmetric at the level of the frequency of photons falling inside and then emerging outside. The effective change of signature will turn the oscillating (complex) exponential into an increasing (real) exponential. This should affect the associated phenomenology. The energy of the emitted photons will not  anymore be the same than the one of the in-falling photons, at least for the so-called  {\it high energy} signal. This component is indeed due to radiation assumed to emerge with an energy equal to the temperature of the Universe at the formation time of the black hole, which cannot be true anymore. This, of course, requires further studies.

\section{Discussion}
\label{sec:conclu}
The apparent change of signature obviously raises some questions, since unlike GR, it is still unclear whether the deformed algebra of constraints really traces a deformation of the algebra of some "hypersurfaces" embedded in a 4-dimensional, signature-changing space. In GR, there is a clear equivalence between the Hamiltonian formulation and the geometrical formulation, while the formalism presented above is solely phrased in an Hamiltonian language. One may however try to reconstruct some notions such as 4-dimensional, {\it effective} world-lines as a way to have some clues on what would be a possible underlying 4-dimensional structure leading to signature change. 

Physically speaking, a geodesic is no more than the world-line followed by some matter content in a given space-time. From a field theoretic viewpoint, geodesics are obtained by taking the optical (or eikonal) limit of the fields equation. By performing this, one can easily recover the fact that photons follow null geodesics from the optical limit of the Maxwell equations in curved space-time \cite{Fleury:2015hgz}, or similarly by starting from the Klein-Gordon equation for a massive scalar field, that massive particles follow time-like geodesics (as a result of the optical limit). If the deformation of the algebra of constraints truly traces a change of signature, this should affect any test fields propagating in the quantum-corrected space-time. An interesting idea would thus be to use the optical limit of test fields propagating in the quantum-corrected space as a way to reconstruct the ``world-lines" followed by massless or massive particles. The starting point being the Hamiltonian framework, the strategy could be the following. Considering e.g. a massive scalar field, or the electromagnetic field, the Hamiltonian and diffeomorphism constraints have to be constructed such as to satisfy the {\it deformed} algebra of constraints, thus ensuring that the dynamics of these test fields is generated by proper representations of the algebra of hypersurface deformation (following the idea that the algebra of constraints of any fields has to represent of the algebra of hypersurface deformations \cite{HOJMAN197688}). From that, one thus obtains the deformed equation of motion for the test fields (in the form of a wave equation), from which one finally implements the optical limit to get the properties of the "world-lines" followed by massive particles, or photons. 

This approach to reconstruct some notions of world-lines is based on the idea that geometrical notions are traced back to the dynamics of some fields, which could be instructive for the underlying 4-dimensional structure (in a very same way that in solid state physics, light propagating in special medium can be described by propagation in a Finsler geometry \cite{Skakala:2008jp}). It may however not be unique. As an exemple, one may argue that the characteristics of the wave equations, as studied in \cite{Bojowald:2015gra}, is another way to derive some notion of effective world-lines. This approach could be interesting not only to shed light on the possible 4-dimensional structure underlying the deformed algebra of constraints, but also in order to search for phenomenological consequences of the change of signature. It is expected from the above procedure to have an equation of motion for point-like particles, which however propagate in a signature-changing space. This could for example serve for studying the behavior of particles traversing e.g. bouncing black holes.

\bibliography{refs}

\begin{thebibliography}{42}
\expandafter\ifx\csname natexlab\endcsname\relax\def\natexlab#1{#1}\fi
\expandafter\ifx\csname bibnamefont\endcsname\relax
  \def\bibnamefont#1{#1}\fi
\expandafter\ifx\csname bibfnamefont\endcsname\relax
  \def\bibfnamefont#1{#1}\fi
\expandafter\ifx\csname citenamefont\endcsname\relax
  \def\citenamefont#1{#1}\fi
\expandafter\ifx\csname url\endcsname\relax
  \def\url#1{\texttt{#1}}\fi
\expandafter\ifx\csname urlprefix\endcsname\relax\def\urlprefix{URL }\fi
\providecommand{\bibinfo}[2]{#2}
\providecommand{\eprint}[2][]{\url{#2}}

\bibitem[{\citenamefont{Arnowitt et~al.}(1959)\citenamefont{Arnowitt, Deser,
  and Misner}}]{PhysRev.116.1322}
\bibinfo{author}{\bibfnamefont{R.}~\bibnamefont{Arnowitt}},
  \bibinfo{author}{\bibfnamefont{S.}~\bibnamefont{Deser}}, \bibnamefont{and}
  \bibinfo{author}{\bibfnamefont{C.~W.} \bibnamefont{Misner}},
  \bibinfo{journal}{Phys. Rev.} \textbf{\bibinfo{volume}{116}},
  \bibinfo{pages}{1322} (\bibinfo{year}{1959}),
  \urlprefix\url{http://link.aps.org/doi/10.1103/PhysRev.116.1322}.

\bibitem[{\citenamefont{Thiemann}(2008)}]{lqg6}
\bibinfo{author}{\bibfnamefont{T.}~\bibnamefont{Thiemann}},
  \emph{\bibinfo{title}{{Modern Canonical Quantum General Relativity}}}
  (\bibinfo{publisher}{Cambridge University Press}, \bibinfo{year}{2008}),
  \bibinfo{note}{iSBN-10:0521741874}.

\bibitem[{\citenamefont{Hojman et~al.}(1976)\citenamefont{Hojman, Kucha\v{r},
  and Teitelboim}}]{HOJMAN197688}
\bibinfo{author}{\bibfnamefont{S.~A.} \bibnamefont{Hojman}},
  \bibinfo{author}{\bibfnamefont{K.}~\bibnamefont{Kucha\v{r}}},
  \bibnamefont{and}
  \bibinfo{author}{\bibfnamefont{C.}~\bibnamefont{Teitelboim}},
  \bibinfo{journal}{Annals of Physics} \textbf{\bibinfo{volume}{96}},
  \bibinfo{pages}{88 } (\bibinfo{year}{1976}), ISSN \bibinfo{issn}{0003-4916},
  \urlprefix\url{http://www.sciencedirect.com/science/article/pii/0003491676901123}.

\bibitem[{\citenamefont{Barrau et~al.}(2014{\natexlab{a}})\citenamefont{Barrau,
  Cailleteau, Grain, and Mielczarek}}]{Barrau:2013ula}
\bibinfo{author}{\bibfnamefont{A.}~\bibnamefont{Barrau}},
  \bibinfo{author}{\bibfnamefont{T.}~\bibnamefont{Cailleteau}},
  \bibinfo{author}{\bibfnamefont{J.}~\bibnamefont{Grain}}, \bibnamefont{and}
  \bibinfo{author}{\bibfnamefont{J.}~\bibnamefont{Mielczarek}},
  \bibinfo{journal}{Class.Quant.Grav.} \textbf{\bibinfo{volume}{31}},
  \bibinfo{pages}{053001} (\bibinfo{year}{2014}{\natexlab{a}}),
  \eprint{1309.6896}.

\bibitem[{\citenamefont{Bojowald et~al.}(2015)\citenamefont{Bojowald, Brahma,
  and Reyes}}]{Bojowald:2015zha}
\bibinfo{author}{\bibfnamefont{M.}~\bibnamefont{Bojowald}},
  \bibinfo{author}{\bibfnamefont{S.}~\bibnamefont{Brahma}}, \bibnamefont{and}
  \bibinfo{author}{\bibfnamefont{J.~D.} \bibnamefont{Reyes}},
  \bibinfo{journal}{Phys. Rev.} \textbf{\bibinfo{volume}{D92}},
  \bibinfo{pages}{045043} (\bibinfo{year}{2015}), \eprint{1507.00329}.

\bibitem[{\citenamefont{Bojowald and Brahma}(2015)}]{Bojowald:2015sta}
\bibinfo{author}{\bibfnamefont{M.}~\bibnamefont{Bojowald}} \bibnamefont{and}
  \bibinfo{author}{\bibfnamefont{S.}~\bibnamefont{Brahma}},
  \bibinfo{journal}{Phys. Rev.} \textbf{\bibinfo{volume}{D92}},
  \bibinfo{pages}{065002} (\bibinfo{year}{2015}), \eprint{1507.00679}.

\bibitem[{\citenamefont{Barrau et~al.}(2015)\citenamefont{Barrau, Bojowald,
  Calcagni, Grain, and Kagan}}]{eucl3}
\bibinfo{author}{\bibfnamefont{A.}~\bibnamefont{Barrau}},
  \bibinfo{author}{\bibfnamefont{M.}~\bibnamefont{Bojowald}},
  \bibinfo{author}{\bibfnamefont{G.}~\bibnamefont{Calcagni}},
  \bibinfo{author}{\bibfnamefont{J.}~\bibnamefont{Grain}}, \bibnamefont{and}
  \bibinfo{author}{\bibfnamefont{M.}~\bibnamefont{Kagan}},
  \bibinfo{journal}{JCAP} \textbf{\bibinfo{volume}{1505}}, \bibinfo{pages}{051}
  (\bibinfo{year}{2015}), \eprint{1404.1018}.

\bibitem[{\citenamefont{Rovelli}(2011)}]{lqg3}
\bibinfo{author}{\bibfnamefont{C.}~\bibnamefont{Rovelli}},
  \bibinfo{journal}{PoS} \textbf{\bibinfo{volume}{QGQGS2011}},
  \bibinfo{pages}{003} (\bibinfo{year}{2011}), \eprint{1102.3660}.

\bibitem[{\citenamefont{Bianchi et~al.}(2010)\citenamefont{Bianchi, Rovelli,
  and Vidotto}}]{Bianchi:2010zs}
\bibinfo{author}{\bibfnamefont{E.}~\bibnamefont{Bianchi}},
  \bibinfo{author}{\bibfnamefont{C.}~\bibnamefont{Rovelli}}, \bibnamefont{and}
  \bibinfo{author}{\bibfnamefont{F.}~\bibnamefont{Vidotto}},
  \bibinfo{journal}{Phys. Rev.} \textbf{\bibinfo{volume}{D82}},
  \bibinfo{pages}{084035} (\bibinfo{year}{2010}), \eprint{1003.3483}.

\bibitem[{\citenamefont{Agullo et~al.}(2012)\citenamefont{Agullo, Ashtekar, and
  Nelson}}]{agullo2}
\bibinfo{author}{\bibfnamefont{I.}~\bibnamefont{Agullo}},
  \bibinfo{author}{\bibfnamefont{A.}~\bibnamefont{Ashtekar}}, \bibnamefont{and}
  \bibinfo{author}{\bibfnamefont{W.}~\bibnamefont{Nelson}},
  \bibinfo{journal}{Phys. Rev. Lett.} \textbf{\bibinfo{volume}{109}},
  \bibinfo{pages}{251301} (\bibinfo{year}{2012}), \eprint{1209.1609}.

\bibitem[{\citenamefont{Cailleteau
  et~al.}(2012{\natexlab{a}})\citenamefont{Cailleteau, Barrau, Grain, and
  Vidotto}}]{eucl2}
\bibinfo{author}{\bibfnamefont{T.}~\bibnamefont{Cailleteau}},
  \bibinfo{author}{\bibfnamefont{A.}~\bibnamefont{Barrau}},
  \bibinfo{author}{\bibfnamefont{J.}~\bibnamefont{Grain}}, \bibnamefont{and}
  \bibinfo{author}{\bibfnamefont{F.}~\bibnamefont{Vidotto}},
  \bibinfo{journal}{Phys. Rev.} \textbf{\bibinfo{volume}{D86}},
  \bibinfo{pages}{087301} (\bibinfo{year}{2012}{\natexlab{a}}),
  \eprint{1206.6736}.

\bibitem[{\citenamefont{Langlois}(1994)}]{Langlois:1994ec}
\bibinfo{author}{\bibfnamefont{D.}~\bibnamefont{Langlois}},
  \bibinfo{journal}{Class. Quant. Grav.} \textbf{\bibinfo{volume}{11}},
  \bibinfo{pages}{389} (\bibinfo{year}{1994}).

\bibitem[{\citenamefont{Cailleteau and Barrau}(2012)}]{Cailleteau:2011mi}
\bibinfo{author}{\bibfnamefont{T.}~\bibnamefont{Cailleteau}} \bibnamefont{and}
  \bibinfo{author}{\bibfnamefont{A.}~\bibnamefont{Barrau}},
  \bibinfo{journal}{Phys. Rev.} \textbf{\bibinfo{volume}{D85}},
  \bibinfo{pages}{123534} (\bibinfo{year}{2012}), \eprint{1111.7192}.

\bibitem[{\citenamefont{Ashtekar and Singh}(2011)}]{lqc9}
\bibinfo{author}{\bibfnamefont{A.}~\bibnamefont{Ashtekar}} \bibnamefont{and}
  \bibinfo{author}{\bibfnamefont{P.}~\bibnamefont{Singh}},
  \bibinfo{journal}{Class. Quant. Grav.} \textbf{\bibinfo{volume}{28}},
  \bibinfo{pages}{213001} (\bibinfo{year}{2011}), \eprint{1108.0893}.

\bibitem[{\citenamefont{Bojowald}(2008)}]{lqc10}
\bibinfo{author}{\bibfnamefont{M.}~\bibnamefont{Bojowald}},
  \bibinfo{journal}{Living Reviews in Relativity} \textbf{\bibinfo{volume}{11}}
  (\bibinfo{year}{2008}),
  \bibinfo{note}{\href{http://www.livingreviews.org/lrr-2008-4}{http://www.livingreviews.org/lrr-2008-4}}.

\bibitem[{\citenamefont{Grain}(2016)}]{Grain:2016jlq}
\bibinfo{author}{\bibfnamefont{J.}~\bibnamefont{Grain}} (\bibinfo{year}{2016}),
  \eprint{1606.03271}.

\bibitem[{\citenamefont{Agullo et~al.}(2013{\natexlab{a}})\citenamefont{Agullo,
  Ashtekar, and Nelson}}]{agullo1}
\bibinfo{author}{\bibfnamefont{I.}~\bibnamefont{Agullo}},
  \bibinfo{author}{\bibfnamefont{A.}~\bibnamefont{Ashtekar}}, \bibnamefont{and}
  \bibinfo{author}{\bibfnamefont{W.}~\bibnamefont{Nelson}},
  \bibinfo{journal}{Class.Quant.Grav.} \textbf{\bibinfo{volume}{30}},
  \bibinfo{pages}{085014} (\bibinfo{year}{2013}{\natexlab{a}}),
  \eprint{1302.0254}.

\bibitem[{\citenamefont{Agullo et~al.}(2013{\natexlab{b}})\citenamefont{Agullo,
  Ashtekar, and Nelson}}]{agullo3}
\bibinfo{author}{\bibfnamefont{I.}~\bibnamefont{Agullo}},
  \bibinfo{author}{\bibfnamefont{A.}~\bibnamefont{Ashtekar}}, \bibnamefont{and}
  \bibinfo{author}{\bibfnamefont{W.}~\bibnamefont{Nelson}},
  \bibinfo{journal}{Phys. Rev.} \textbf{\bibinfo{volume}{D87}},
  \bibinfo{pages}{043507} (\bibinfo{year}{2013}{\natexlab{b}}),
  \eprint{1211.1354}.

\bibitem[{\citenamefont{Barrau and Bolliet}(2016)}]{Barrau:2016nwy}
\bibinfo{author}{\bibfnamefont{A.}~\bibnamefont{Barrau}} \bibnamefont{and}
  \bibinfo{author}{\bibfnamefont{B.}~\bibnamefont{Bolliet}}
  (\bibinfo{year}{2016}), \eprint{1602.04452}.

\bibitem[{\citenamefont{Mielczarek et~al.}(2012)\citenamefont{Mielczarek,
  Cailleteau, Barrau, and Grain}}]{tom1}
\bibinfo{author}{\bibfnamefont{J.}~\bibnamefont{Mielczarek}},
  \bibinfo{author}{\bibfnamefont{T.}~\bibnamefont{Cailleteau}},
  \bibinfo{author}{\bibfnamefont{A.}~\bibnamefont{Barrau}}, \bibnamefont{and}
  \bibinfo{author}{\bibfnamefont{J.}~\bibnamefont{Grain}},
  \bibinfo{journal}{Class. Quant. Grav.} \textbf{\bibinfo{volume}{29}},
  \bibinfo{pages}{085009} (\bibinfo{year}{2012}), \eprint{1106.3744}.

\bibitem[{\citenamefont{Cailleteau
  et~al.}(2012{\natexlab{b}})\citenamefont{Cailleteau, Mielczarek, Barrau, and
  Grain}}]{tom2}
\bibinfo{author}{\bibfnamefont{T.}~\bibnamefont{Cailleteau}},
  \bibinfo{author}{\bibfnamefont{J.}~\bibnamefont{Mielczarek}},
  \bibinfo{author}{\bibfnamefont{A.}~\bibnamefont{Barrau}}, \bibnamefont{and}
  \bibinfo{author}{\bibfnamefont{J.}~\bibnamefont{Grain}},
  \bibinfo{journal}{Class. Quant. Grav.} \textbf{\bibinfo{volume}{29}},
  \bibinfo{pages}{095010} (\bibinfo{year}{2012}{\natexlab{b}}),
  \eprint{1111.3535}.

\bibitem[{\citenamefont{Wilson-Ewing}(2012{\natexlab{a}})}]{WilsonEwing:2011es}
\bibinfo{author}{\bibfnamefont{E.}~\bibnamefont{Wilson-Ewing}},
  \bibinfo{journal}{Class. Quant. Grav.} \textbf{\bibinfo{volume}{29}},
  \bibinfo{pages}{085005} (\bibinfo{year}{2012}{\natexlab{a}}),
  \eprint{1108.6265}.

\bibitem[{\citenamefont{Wilson-Ewing}(2012{\natexlab{b}})}]{ed}
\bibinfo{author}{\bibfnamefont{E.}~\bibnamefont{Wilson-Ewing}},
  \bibinfo{journal}{Class. Quant. Grav.} \textbf{\bibinfo{volume}{29}},
  \bibinfo{pages}{215013} (\bibinfo{year}{2012}{\natexlab{b}}),
  \eprint{1205.3370}.

\bibitem[{\citenamefont{Linsefors et~al.}(2013)\citenamefont{Linsefors,
  Cailleteau, Barrau, and Grain}}]{lcbg}
\bibinfo{author}{\bibfnamefont{L.}~\bibnamefont{Linsefors}},
  \bibinfo{author}{\bibfnamefont{T.}~\bibnamefont{Cailleteau}},
  \bibinfo{author}{\bibfnamefont{A.}~\bibnamefont{Barrau}}, \bibnamefont{and}
  \bibinfo{author}{\bibfnamefont{J.}~\bibnamefont{Grain}},
  \bibinfo{journal}{Phys. Rev.} \textbf{\bibinfo{volume}{D87}},
  \bibinfo{pages}{107503} (\bibinfo{year}{2013}), \eprint{1212.2852}.

\bibitem[{\citenamefont{Bolliet et~al.}(2015)\citenamefont{Bolliet, Grain,
  Stahl, Linsefors, and Barrau}}]{Bolliet:2015bka}
\bibinfo{author}{\bibfnamefont{B.}~\bibnamefont{Bolliet}},
  \bibinfo{author}{\bibfnamefont{J.}~\bibnamefont{Grain}},
  \bibinfo{author}{\bibfnamefont{C.}~\bibnamefont{Stahl}},
  \bibinfo{author}{\bibfnamefont{L.}~\bibnamefont{Linsefors}},
  \bibnamefont{and} \bibinfo{author}{\bibfnamefont{A.}~\bibnamefont{Barrau}},
  \bibinfo{journal}{Phys.Rev.} \textbf{\bibinfo{volume}{D91}},
  \bibinfo{pages}{084035} (\bibinfo{year}{2015}), \eprint{1502.02431}.

\bibitem[{\citenamefont{Schander et~al.}(2016)\citenamefont{Schander, Barrau,
  Bolliet, Linsefors, Mielczarek, and Grain}}]{Schander:2015eja}
\bibinfo{author}{\bibfnamefont{S.}~\bibnamefont{Schander}},
  \bibinfo{author}{\bibfnamefont{A.}~\bibnamefont{Barrau}},
  \bibinfo{author}{\bibfnamefont{B.}~\bibnamefont{Bolliet}},
  \bibinfo{author}{\bibfnamefont{L.}~\bibnamefont{Linsefors}},
  \bibinfo{author}{\bibfnamefont{J.}~\bibnamefont{Mielczarek}},
  \bibnamefont{and} \bibinfo{author}{\bibfnamefont{J.}~\bibnamefont{Grain}},
  \bibinfo{journal}{Phys. Rev.} \textbf{\bibinfo{volume}{D93}},
  \bibinfo{pages}{023531} (\bibinfo{year}{2016}), \eprint{1508.06786}.

\bibitem[{\citenamefont{Bolliet et~al.}(2016)\citenamefont{Bolliet, Barrau,
  Grain, and Schander}}]{Bolliet:2015raa}
\bibinfo{author}{\bibfnamefont{B.}~\bibnamefont{Bolliet}},
  \bibinfo{author}{\bibfnamefont{A.}~\bibnamefont{Barrau}},
  \bibinfo{author}{\bibfnamefont{J.}~\bibnamefont{Grain}}, \bibnamefont{and}
  \bibinfo{author}{\bibfnamefont{S.}~\bibnamefont{Schander}},
  \bibinfo{journal}{Phys. Rev.} \textbf{\bibinfo{volume}{D93}},
  \bibinfo{pages}{124011} (\bibinfo{year}{2016}), \eprint{1510.08766}.

\bibitem[{\citenamefont{Mielczarek et~al.}(2014)\citenamefont{Mielczarek,
  Linsefors, and Barrau}}]{Mielczarek:2014kea}
\bibinfo{author}{\bibfnamefont{J.}~\bibnamefont{Mielczarek}},
  \bibinfo{author}{\bibfnamefont{L.}~\bibnamefont{Linsefors}},
  \bibnamefont{and} \bibinfo{author}{\bibfnamefont{A.}~\bibnamefont{Barrau}}
  (\bibinfo{year}{2014}), \eprint{1411.0272}.

\bibitem[{\citenamefont{Mielczarek}(2014{\natexlab{a}})}]{Mielczarek:2013xaa}
\bibinfo{author}{\bibfnamefont{J.}~\bibnamefont{Mielczarek}},
  \bibinfo{journal}{JCAP} \textbf{\bibinfo{volume}{1403}}, \bibinfo{pages}{048}
  (\bibinfo{year}{2014}{\natexlab{a}}), \eprint{1311.1344}.

\bibitem[{\citenamefont{Bojowald and Paily}(2013)}]{Bojowald:2012ux}
\bibinfo{author}{\bibfnamefont{M.}~\bibnamefont{Bojowald}} \bibnamefont{and}
  \bibinfo{author}{\bibfnamefont{G.~M.} \bibnamefont{Paily}},
  \bibinfo{journal}{Phys. Rev.} \textbf{\bibinfo{volume}{D87}},
  \bibinfo{pages}{044044} (\bibinfo{year}{2013}), \eprint{1212.4773}.

\bibitem[{\citenamefont{Mielczarek}(2014{\natexlab{b}})}]{Mielczarek:2013rva}
\bibinfo{author}{\bibfnamefont{J.}~\bibnamefont{Mielczarek}},
  \bibinfo{journal}{Europhys. Lett.} \textbf{\bibinfo{volume}{108}},
  \bibinfo{pages}{40003} (\bibinfo{year}{2014}{\natexlab{b}}),
  \eprint{1304.2208}.

\bibitem[{\citenamefont{Bojowald and Mielczarek}(2015)}]{Bojowald:2015gra}
\bibinfo{author}{\bibfnamefont{M.}~\bibnamefont{Bojowald}} \bibnamefont{and}
  \bibinfo{author}{\bibfnamefont{J.}~\bibnamefont{Mielczarek}},
  \bibinfo{journal}{JCAP} \textbf{\bibinfo{volume}{1508}}, \bibinfo{pages}{052}
  (\bibinfo{year}{2015}), \eprint{1503.09154}.

\bibitem[{\citenamefont{Bojowald}(2015)}]{Bojowald:2014zla}
\bibinfo{author}{\bibfnamefont{M.}~\bibnamefont{Bojowald}},
  \bibinfo{journal}{Front.in Phys.} \textbf{\bibinfo{volume}{3}},
  \bibinfo{pages}{33} (\bibinfo{year}{2015}), \eprint{1409.3157}.

\bibitem[{\citenamefont{Rovelli and Vidotto}(2014)}]{Rovelli:2014cta}
\bibinfo{author}{\bibfnamefont{C.}~\bibnamefont{Rovelli}} \bibnamefont{and}
  \bibinfo{author}{\bibfnamefont{F.}~\bibnamefont{Vidotto}},
  \bibinfo{journal}{Int. J. Mod. Phys.} \textbf{\bibinfo{volume}{D23}},
  \bibinfo{pages}{1442026} (\bibinfo{year}{2014}), \eprint{1401.6562}.

\bibitem[{\citenamefont{Haggard and Rovelli}(2015)}]{Haggard:2014rza}
\bibinfo{author}{\bibfnamefont{H.~M.} \bibnamefont{Haggard}} \bibnamefont{and}
  \bibinfo{author}{\bibfnamefont{C.}~\bibnamefont{Rovelli}},
  \bibinfo{journal}{Phys. Rev.} \textbf{\bibinfo{volume}{D92}},
  \bibinfo{pages}{104020} (\bibinfo{year}{2015}), \eprint{1407.0989}.

\bibitem[{\citenamefont{Barrau and Rovelli}(2014)}]{Barrau:2014hda}
\bibinfo{author}{\bibfnamefont{A.}~\bibnamefont{Barrau}} \bibnamefont{and}
  \bibinfo{author}{\bibfnamefont{C.}~\bibnamefont{Rovelli}},
  \bibinfo{journal}{Phys. Lett.} \textbf{\bibinfo{volume}{B739}},
  \bibinfo{pages}{405} (\bibinfo{year}{2014}), \eprint{1404.5821}.

\bibitem[{\citenamefont{Barrau et~al.}(2014{\natexlab{b}})\citenamefont{Barrau,
  Rovelli, and Vidotto}}]{Barrau:2014yka}
\bibinfo{author}{\bibfnamefont{A.}~\bibnamefont{Barrau}},
  \bibinfo{author}{\bibfnamefont{C.}~\bibnamefont{Rovelli}}, \bibnamefont{and}
  \bibinfo{author}{\bibfnamefont{F.}~\bibnamefont{Vidotto}},
  \bibinfo{journal}{Phys. Rev.} \textbf{\bibinfo{volume}{D90}},
  \bibinfo{pages}{127503} (\bibinfo{year}{2014}{\natexlab{b}}),
  \eprint{1409.4031}.

\bibitem[{\citenamefont{Barrau et~al.}(2016{\natexlab{a}})\citenamefont{Barrau,
  Bolliet, Vidotto, and Weimer}}]{Barrau:2015uca}
\bibinfo{author}{\bibfnamefont{A.}~\bibnamefont{Barrau}},
  \bibinfo{author}{\bibfnamefont{B.}~\bibnamefont{Bolliet}},
  \bibinfo{author}{\bibfnamefont{F.}~\bibnamefont{Vidotto}}, \bibnamefont{and}
  \bibinfo{author}{\bibfnamefont{C.}~\bibnamefont{Weimer}},
  \bibinfo{journal}{JCAP} \textbf{\bibinfo{volume}{1602}}, \bibinfo{pages}{022}
  (\bibinfo{year}{2016}{\natexlab{a}}), \eprint{1507.05424}.

\bibitem[{\citenamefont{Barrau et~al.}(2016{\natexlab{b}})\citenamefont{Barrau,
  Bolliet, Schutten, and Vidotto}}]{Barrau:2016fcg}
\bibinfo{author}{\bibfnamefont{A.}~\bibnamefont{Barrau}},
  \bibinfo{author}{\bibfnamefont{B.}~\bibnamefont{Bolliet}},
  \bibinfo{author}{\bibfnamefont{M.}~\bibnamefont{Schutten}}, \bibnamefont{and}
  \bibinfo{author}{\bibfnamefont{F.}~\bibnamefont{Vidotto}}
  (\bibinfo{year}{2016}{\natexlab{b}}), \eprint{1606.08031}.

\bibitem[{\citenamefont{Ashtekar and Bojowald}(2005)}]{Ashtekar:2005cj}
\bibinfo{author}{\bibfnamefont{A.}~\bibnamefont{Ashtekar}} \bibnamefont{and}
  \bibinfo{author}{\bibfnamefont{M.}~\bibnamefont{Bojowald}},
  \bibinfo{journal}{Class. Quant. Grav.} \textbf{\bibinfo{volume}{22}},
  \bibinfo{pages}{3349} (\bibinfo{year}{2005}), \eprint{gr-qc/0504029}.

\bibitem[{\citenamefont{Fleury}(2015)}]{Fleury:2015hgz}
\bibinfo{author}{\bibfnamefont{P.}~\bibnamefont{Fleury}}, Ph.D. thesis,
  \bibinfo{school}{Paris, Inst. Astrophys.} (\bibinfo{year}{2015}),
  \eprint{1511.03702},
  \urlprefix\url{http://inspirehep.net/record/1404129/files/arXiv:1511.03702.pdf}.

\bibitem[{\citenamefont{Skakala and Visser}(2010)}]{Skakala:2008jp}
\bibinfo{author}{\bibfnamefont{J.}~\bibnamefont{Skakala}} \bibnamefont{and}
  \bibinfo{author}{\bibfnamefont{M.}~\bibnamefont{Visser}},
  \bibinfo{journal}{Int. J. Mod. Phys.} \textbf{\bibinfo{volume}{D19}},
  \bibinfo{pages}{1119} (\bibinfo{year}{2010}), \eprint{0806.0950}.

\end{thebibliography}

\end{document}